\documentclass[useAMS,usenatbib]{mn2e}
\usepackage{psfig, epsf, epsfig}
\title[The origin of BCDs]{Formation of blue compact dwarf
galaxies from  merging and interacting gas-rich dwarfs}
\author[K. Bekki]
{Kenji Bekki${}^1$\thanks{E-mail:
bekki@phys.unsw.edu.au} \\
       ${}^1$School of Physics, University of New South Wales,
              Sydney 2052, NSW, Australia}

\begin{document}

\date{Accepted, Received 2005 February 20; in original form }

\pagerange{\pageref{firstpage}--\pageref{lastpage}} \pubyear{2005}

\maketitle

\label{firstpage}

\begin{abstract}

We  present  the results of numerical simulations which show
the formation of blue compact dwarf (BCD) galaxies from
merging between very gas-rich dwarfs with extended HI gas disks. 
We show that dwarf-dwarf merging can trigger central starbursts
and form massive compact cores dominated by young stellar populations.
We also show that the pre-existing old stellar components
in merger precursor dwarfs can become diffuse low surface brightness
components after merging.
The compact cores dominated by  younger  stellar populations
and embedded in more diffusely distributed older ones
can be morphologically classified as BCDs.
Since new stars can be formed from gas transferred from the outer 
part of the extended gas disks of merger precursors, 
new stars can be very metal-poor (${\rm [Fe/H]} < -1$). 
Owing to very high gaseous  pressure
exceeding  10$^5$ $k_B$ (where $k_B$ is the Boltzmann
constant) during merging, 
compact star clusters can be formed in forming BCDs.
The BCDs formed from  merging can still have
extended HI gas disks surrounding their blue compact cores.
We discuss whether tidal interaction of gas-rich dwarfs
without merging can also form  BCDs.

\end{abstract}


\section{Introduction}

Blue compact dwarf galaxies (BCDs)
are defined as dwarfs  having $M_{\rm B} \ge -18$ mag, 
optical sizes smaller than 1 kpc, and spectra similar to HII regions
of spiral galaxies
(Thuan \& Martin 1981)
and their formation and evolution processes 
have long been discussed by many authors.
Some gas-rich BCDs dominated by very metal-poor, young stellar
populations (e.g., I Zw 18)  were once suggested to 
be  undergoing their first starbursts 
(e.g., Searle et al. 1973; Aloisi et al. 1999; Thuan et al. 1999),
recent observations have however confirmed that  most of BCDs have 
underlying older stellar populations at least a few $10^9$ yr old
(e.g., James 1994; Cair\'os et al. 2001; Amor\'in et al. 2007).
Although optical structures
(Gil de Paz et al. 2003),  
HI properties (van Zee et al. 1998),
and chemical abundances (e.g., Hunter \& Hoffman 1999)
in BCDs have been discussed in terms of evolutionary
links between BCDs and other types of dwarfs such as dwarf
spheroidal (dSphs),  ellipticals (dEs), and irregulars (dIrrs),
physical connections  between these dwarfs  remain unclear.

\begin{table*}
\centering
\begin{minipage}{185mm}
\caption{Model parameters for N-body simulations}
\begin{tabular}{ccccccccccc}
Model
& {Morphology
\footnote{The initial morphology of old stars: ``disk'' and ``spheroid''
represent the dwarf disk and spheroidal models, respectively.}}
& {$M_{\rm s}$
\footnote{The initial total stellar  mass of a dwarf in
units of  $10^{9}$  ${\rm M_{\odot}}$.}}
& {$r_{\rm s}$
\footnote{The initial stellar size of a dwarf in
units of  kpc.}}
& {$M_{\rm g}$
\footnote{The initial gas mass in units of $10^9 {\rm M}_{\odot}$.}}
& {$s_{\rm g}$
\footnote{The initial size ratio of gas disk to stellar one.}}
& {Orbit types 
\footnote{``PR'' and ``PP' describe  prograde-retrograde
and prograde-prograde orbits, respectively.}}
& {$m_{\rm 2}$
\footnote{The mass ratio of interacting or merging dwarfs.}}
& {$e_{\rm p}$
\footnote{The orbital eccentricity of interacting or merging dwarfs.}}
& {$r_{\rm p}$
\footnote{The pericenter distance of interacting or merging dwarfs
in units of kpc.}} 
& Comments \\
M1 &  disk & 0.4  &  1.25 &  0.8 & 4.0 & PR & 0.5 & 1.0 & 1.0  & merging\\
M2 &  disk & 1.0  &  5.00 &  0.1 & 1.0 & PP & 1.0 & 1.1 & 10.0 &
interaction \\
M3 &  spheroid & 1.0  &  5.00 &  0.1 & 1.0 & PP & 1.0 & 1.1 & 10.0 &
interaction  \\
M4 &  disk & 0.4  &  1.25 &  0.04 & 4.0 & PR & 0.5 & 1.0 & 1.0  & gas-poor\\
\end{tabular}
\end{minipage}
\end{table*}

A growing number of observations have recently suggested
that the origin of  BCDs
can be closely associated with galaxy merging between low-mass dwarfs 
(e.g., Noeske et al. 2001; \"Ostlin et al. 2001).
For example, 
Pustilnik et al. (2001) investigated environments of
BCDs (e.g., the presence or the absence of nearby companion galaxies)
and suggested that at least  80\% of BCDs in the studied sample
have evidence of their starburst components
being triggered either by tidal  interaction or by galaxy merging.
Furthermore, 
observational studies of the six most metal-deficient 
starbursting dwarfs (most of which are BCDs)
by Giant Metrewave Radio Telescope (GMRT) have reported
that they show clear signs of interaction and merging in their HI
properties (Chengalur et al. 2008).  
However,  no theoretical and numerical studies have yet provided
the detailed predictions on
 physical properties of BCDs formed from ``dwarf-dwarf 
merging''.
Therefore, it remain unclear whether  the observed properties
of the BCDs can be consistent  with the merger scenario of BCD formation.

The purpose of this paper is thus to present the results of numerical
simulations which show physical properties of BCDs formed from
dwarf-dwarf merging.
Based on the numerical results,
we discuss (i) the origin of young, metal-poor stellar populations
in BCDs,  (ii) possible evolutionary links between BCDs,
dSphs, dEs, and dIrrs,
and (iii) the roles of  
tidal interaction in the formation of BCDs.

\section{Model}

We investigate chemodynamical evolution of gas-rich mergers
between dwarfs embedded in massive
dark matter halos by using our TREESPH 
chemodynamical codes that combine key ingredients
used in our previous codes:
chemical evolution and supernovae feedback effects
(Bekki \& Shioya 1999),  hydrodynamics 
for strongly self-gravitating gas (Bekki 1997), 
and formation of globular clusters (GCs) and field stars
(Bekki et al. 2002).
This code enables us to investigate chemical and dynamical
evolution of forming BCDs in a self-consistent manner.
We adopt the Burkert profile (Burkert 1995) for the radial
density profile of the dark matter halo of a dwarf,
because it  can be consistent with rotation curve profiles
of dwarfs (Burkert 1995).
The total mass of the dark matter ($M_{\rm dm}$)
for a dwarf
is set to be $8.0 \times 10^9 {\rm M}_{\odot}$
for all models
and the dark matter halo has a large core radius ($r_{\rm c}$)
of 3.5 kpc and the truncation radius of $ 3.4 r_{\rm c}$
(Burkert 1995).
The old stars in the dwarf are assumed to have either
a disky distribution 
(referred to as ``disk model'') or a spheroidal one (``spheroidal
model'').

The stellar component of a dwarf in the disk model
is described as a purely disk  with 
the initial mass of $M_{\rm s}$ and the disk size of $r_{\rm s}$.
The radial ($R$) and vertical ($Z$) density profile
of the initially thin disk  
are  assumed to be
proportional to $\exp (-R/R_{0}) $ with scale length $R_{0}$ = 1 kpc 
and to  ${\rm sech}^2 (Z/Z_{0})$ with scale length $Z_{0}$ = $0.2R_{0}$,
respectively. 
For the spheroidal models,
the stellar spheroids are assumed to have 
the {\it projected} radial density profiles exactly the same
as those for the stellar components
in the above disk models.
In both disk and spheroidal models,
a dwarf has a  thin gas disk and
the gaseous disk is assumed to have the same 
scale length as that of the stellar one.
The size- and mass-ratios of the gaseous disk to the stellar one
are set to be free parameters 
and described as $s_{\rm g}$ and $f_{\rm g}$, respectively:
the size ($r_{\rm g}$) and mass ($M_{\rm g}$)
of the gas disk are $s_{\rm g} r_{\rm s}$
and $f_{\rm g} M_{\rm s}$, respectively.
The HI diameters of gas-rich galaxies are generally observed to be
larger
than their optical disks (Broeils \& van Woerden 1994). A small fraction
of low luminosity galaxies have HI gas envelopes extending out to 4--7
$r_{\rm s}$ (e.g., Hunter 1997) with HI mass to light ratios up to $\sim
20 {\rm M_{\odot}\,L_{\odot}^{-1}}$ (Warren et al. 2004).
Guided by these  observations,  we investigate models 
with $1 \le s_{\rm g} \le 4$ and $0.1 \le f_{\rm g} \le 2$.

Star formation
is modeled by converting  the collisional
gas particles
into  collisionless new stellar particles according to the algorithm
of star formation  described below.
We adopt the Schmidt law (Schmidt 1959)
with exponent $\gamma$ = 1.5 (1.0  $ < $  $\gamma$
$ < $ 2.0, Kennicutt 1998) as the controlling
parameter of the rate of star formation.
The stars formed from gas are called ``new stars'' 
whereas stars initially within a disk  are called ``old stars''
throughout this paper.
Chemical enrichment through star formation and supernova feedback
is assumed to proceed both locally and instantaneously in the present study.
The values of chemical yield and return parameter
are 0.002 and 0.3, respectively,
and the initial gaseous  metallicity 
is ${\rm [Fe/H]} = -1.6$.

Nearby BCDs (e.g., NGC 1705) are observed to have compact young star clusters
(SCs) that can finally evolve into GCs (e.g., Meurer 1993).
We try to investigate whether BCDs formed from dwarf-dwarf merging
and interaction can have SCs. 
We adopt a plausible assumption that 
if gas pressure ($P_{\rm gas}$) 
can exceed a threshold pressure ($P_{\rm th}$),
such high pressure of ISM
can induce the global collapse of 
giant molecular clouds to form massive compact star clusters 
corresponding to SCs 
(Jog \& Solomon 1992; Elmegreen \& Efremov
1997).  
We show the results of the models with 
$P_{\rm th} = 10^5$ $k_{\rm B}$, where $k_{\rm
B}$ is Boltzmann's constant.
The present results do not depend so strongly
on  $P_{\rm th}$
for  $P_{\rm th} = 10^5-10^6 k_{\rm B}$.
New stars formed from gas with 
$P_{\rm gas}$ $>$ $10^5$ $k_{\rm B}$
are identified as SCs rather than field stars in the present study.
We mainly investigate mass fractions of SCs among all new stars for
each model.

The mass ratio of the two merging or interacting dwarfs ($m_2$), 
the pericenter distance ($r_{\rm p}$),
and the eccentricity ($e_{\rm p}$) are 
assumed to be free parameters.
The orbit of the two dwarfs is
set to be
the $xy$ plane and the distance between
the center of mass of the two dIrrs
is  20 kpc.
The spin of each galaxy in a merger
is specified by two angles $\theta_{i}$ and
$\phi_{i}$, where suffix  $i$ is used to identify each galaxy.
$\theta_{i}$ is the angle between the $z$ axis and the vector of
the angular momentum of a disk.
$\phi_{i}$ is the azimuthal angle measured from the $x$ axis to
the projection of the angular momentum vector of a disk onto the $xy$
plane.

Although we run many models with different parameters
with different $m_{2}$, $f_{\rm g}$, $s_{\rm g}$, and orbits of
mergers,
we show only four  representative models 
(M1, M2, M3, and M4) with two different orbital
configurations: the highly inclined prograde-retrograde (``PR'') orbit with
$r_{\rm p}=1.0$ kpc, $e_{\rm p}=1.0$, 
$\theta_{1}=30^{\circ}$,
$\theta_{2}=120^{\circ}$,  $\phi_{1}=90^{\circ}$,
$\phi_{2}=30^{\circ}$
and the prograde-prograde (``PP'')  one with
$r_{\rm p}=10.0$ kpc, $e_{\rm p}=1.1$, 
$\theta_{1}=0^{\circ}$,
$\theta_{2}=0^{\circ}$,  $\phi_{1}=0^{\circ}$,
$\phi_{2}=0^{\circ}$.
We show the results of the model M1 in detail, because it shows typical
behaviors of BCD formation from dwarf-dwarf merging
in the present simulation.
We also show the results of the model M2 and M3 in  which two dwarfs are 
tidally interacting with each other {\it without merging},
because this comparison of the two models enables us to
clarify how interacting galaxies can become BCDs.
The  parameter values  for the four  models are shown in the Table 1.
The total particle number
used in a major merger  model 
is
220000 for collisionless particles
and 30000 for collisional ones. The fixed gravitational
softening length for dark matter, old and new stars, and gas 
are set to be 0.35 kpc, 0.04 kpc, and 0.135 kpc, respectively. 
We confirm that stellar and gaseous disks  are stable
in isolated models (i.e., with no merging and tidal interaction)
even if gas mass fraction are quite large: this is due to the
very extended gas disks in the present models.

\begin{figure}
\psfig{file=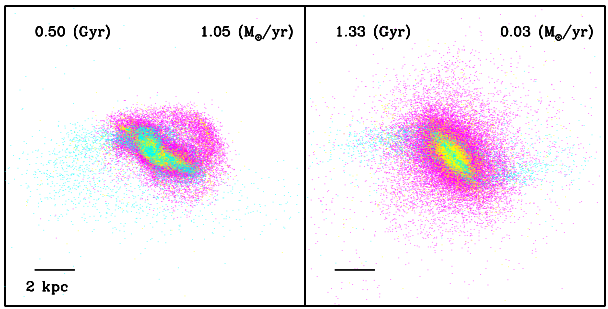,width=8.0cm}
\caption{
Distributions of old stars (magenta), gas (cyan), and new stars
(yellow) projected onto the $x$-$z$ plane 
at $T=0.50$ Gyr (left) and $T=1.33$ Gyr  (right)
in the model M1. 
The time $T$  shown in the upper left corner
for each panel represents the time that has elapsed since
the simulation starts.
The numbers shown in the upper  right
corner for each panel represents
the star formation rate (${\rm M}_{\odot}$ yr$^{-1}$).
The bar in the lower left corner for each panel
measures 2 kpc.
}
\label{Figure. 1}
\end{figure}

\begin{figure}
\psfig{file=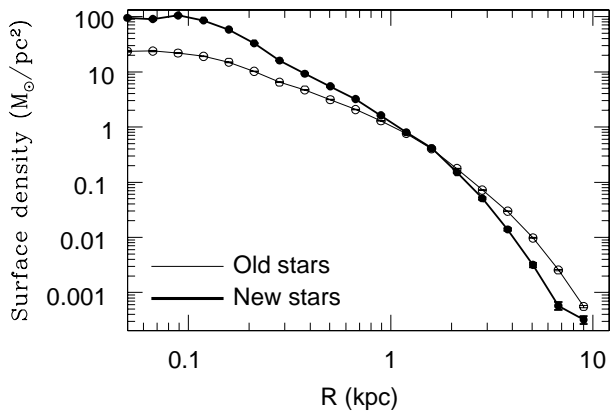,width=8.0cm}
\caption{
The final projected radial density profiles
of the simulated BCD  for old stars (thin)
and new ones (thick) in the model M1.
The profiles 
of the merger remnant well dynamically relaxed are shown here.
}
\label{Figure. 2}
\end{figure}

\begin{figure}
\psfig{file=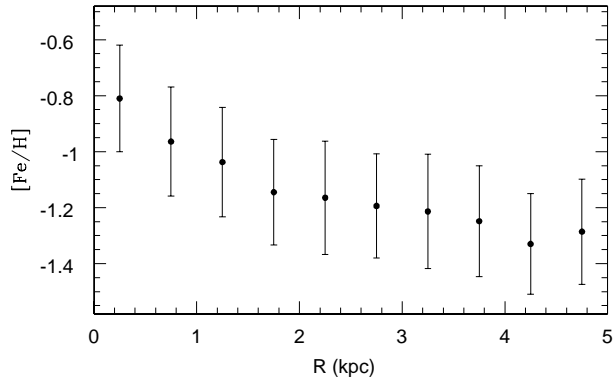,width=8.0cm}
\caption{
The final radial profile of mean metallicities for  new stars 
within the central 5kpc of the merger remnant in the model M1.
The $1 \sigma$ dispersion in the metallicities
for each radial bin is shown by an error bar.
The mean metallicity ($z_{\rm m}$) of new stars 
is ${\rm [Fe/H]} = -0.93$ for this model and $z_{\rm m}$ depends
totally on the adopted initial gaseous metallicities and chemical
yield parameters in the present models. 
}
\label{Figure. 3}
\end{figure}

\section{Results}

Fig. 1 shows mass distributions of different components
(old stars, gas,  and new stars) in the merging model M1 at
two different 
epochs with  star formation rate (SFR) of $1.05 {\rm M}_{\odot}$
yr$^{-1}$
and  $0.03 {\rm M}_{\odot}$ yr$^{-1}$.
SFR reaches its maximum value of $4.13  {\rm M}_{\odot}$ yr$^{-1}$
and about 80\% of initial gas mass is consumed within 1.3 Gyr.
In the late stage of merging
between two gas-rich dwarfs with extended gas disks
($T=0.50$ Gyr),
nuclear starbursts can be triggered
owing to efficient inward transfer of gas 
from the outer HI disks  of the merger progenitor dwarfs. 
As a result of nuclear starbursts,
the compact core 
dominated by young  stars
can be formed and embedded in old stars and gas.

The spatial distribution of old stars 
clearly shows a disturbed morphology 
in the outer part of the merger ($T=0.50$ Gyr).
If the stellar population synthesis models
for metallicities of ${\rm [Fe/H]} = -1.28$ and ages of 0.5 Gyr 
by Vazdekis et al. (1996) are  applied to the central
core dominated by young stars,
then $B-R$ is estimated to be $\sim 0.6$,
which is consistent with the observed range of  $B-R$ in BCDs
(Gil De Paz et al. 2003).
This  merger with a  blue compact  core dominated by young stars
and an outer disturbed morphology 
can be thus identified as a BCD.

Fig. 1 clearly shows that after strong starbursts,
the dwarf-dwarf merger can soon get dynamically relaxed
to form a flattened spheroid with a central compact core
and a HI envelope ($T=1.33$ Gyr).
The HI envelope with the mass of $1.6 \times 10^8 {\rm M}_{\odot}$
(i.e., $M_{\rm g}/M_{\rm s}=0.27$)
has a disky distribution with its outer part
strongly warped. The SFR after strong starbursts can still become
relatively high (e.g., $\sim 0.1 {\rm M}_{\odot}$ yr$^{-1}$ at $T=1.2$ Gyr) 
in a sporadic way
owing to the presence of the  gas disk. 
This  merger remnant 
with a compact core,
a more regular morphology,
and a higher SFR 
therefore can be also identified as a BCD.

The projected radial density profiles of old and new stars
shown in Fig. 2
clearly demonstrate that the merger remnant is dominated
by new stars formed from nuclear starbursts in the central
1 kpc.
If we adopt
$M_{\rm s}/L_{\rm B}$ (i.e., ratio of stellar mass to light in $B$-band) of 1.0,
$B$-band surface brightness (${\mu}_{\rm B}$) distribution
of old (new) stars
ranges from $\sim 24$ ($\sim 22$) mag arcsec$^{-2}$
at $R\sim 0.1$ kpc
to $\sim 29$ ($\sim 29$) mag arcsec$^{-2}$ at $R\sim 2$ kpc
in this model.
Since $M_{\rm s}/L_{\rm B}$ is significantly smaller than 1
for stars younger than 1 Gyr (e.g., Vazdekis et al. 1996),
${\mu}_{\rm B}$ in the central region of the simulated BCD would be
significantly  higher than the above.

Fig. 3 shows that metallicities in ${\rm [Fe/H]}$
for new stars
within the central 5 kpc of the merger remnant
ranges from $\sim -1.3$ to $\sim -0.8$: the simulated BCD
has a significant {\it negative} metallicity gradient
of new stellar populations.
Owing to inhomogeneous  chemical mixing during starbursts,
the metallicity dispersion at a give radius
becomes  significantly large ($\sim 0.2$ dex).
The derived very small metallicities are due essentially
to the inward transfer of metal-poor gas initially in
the extended HI disks of merger progenitor dwarfs.
It should be however stressed here that 
the metallicities of new stars 
depend strongly on the adopted initial gaseous metallicities
in the present models.
These results imply  that younger stellar populations in BCDs,
which consists of their blue compact cores,
can be very metal-poor (${\rm [Fe/H]} <-1$),
{\it because  gas that formed  the younger populations
originate from chemically  pristine HI gas
initially in the outer part of merger progenitor dwarfs.}

\begin{figure}
\psfig{file=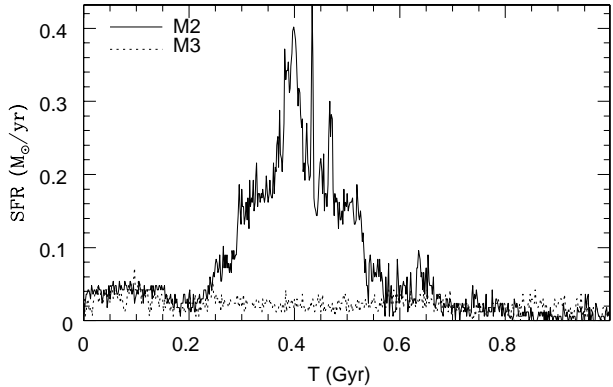,width=8.0cm}
\caption{
The time evolution of star formation rate (SFR) for 
the tidal interaction models M2 (solid) and M3 (dotted). 
}
\label{Figure. 4}
\end{figure}

Fig. 4  shows time evolution of SFR  for the tidal interaction
models M2 and M3 in which the initial  dwarf is modeled as
a disk and a spheroid, respectively.
As shown in Fig. 4,
a secondary starburst can be triggered to 
form a compact core dominated by young stars
only in the disk model (M2):
nuclear starbursts that form blue compact cores
in BCDs
are highly  unlikely to be   triggered in interacting dwarfs,
if the underlying old stars have
spherical distributions rather than disks. 
The interacting dwarf with a  compact core in the model M2
shows an exponential radial density profile 
in its  outer diffuse stellar component
dominated by old stars.
The time scale for which SFR is rather  high 
($>0.1 {\rm M}_{\odot}$  yr$^{-1}$)
in merging and interacting dwarfs is quite short (an order of $\sim 0.1$
Gyr) in the present models. Therefore, the time scale for 
merging and interacting dwarfs to be observationally identified as
BCDs {\it with  very strong emission lines}  can be also short 
(an order of $\sim 0.1$ Gyr).

The present models predict that gaseous pressure during
starbursts triggered by galaxy interaction and merging
can become very high ($P_{\rm gas}$ $>$ $10^5$ $k_{\rm B}$)
so that SCs can be formed in BCDs.
For example, the mass  fraction of  SCs is 0.22 in the model
M1, though not all of these may evolve into GCs finally owing
to internal and external destruction processes after their formation.
This result clearly suggests that BCDs formed from dwarf-dwarf merging
can contain young GCs owing to tidal compression of gas during the  merging. 
Furthermore,
it is highly likely that nuclear star clusters in BCDs  can
be formed from nuclear starbursts during dwarf-dwarf merging.
We thus suggest that the observed presence of young SCs and
blue stellar nuclei in BCDs is possible evidence for
BCD formation via dwarf-dwarf merging.

Lastly, we describe the results of the gas-poor  model M4 in which
a BCD can not be formed. 
Although the SFR can become high 
($\sim 0.6 {\rm M}_{\odot}$ yr$^{-1}$) in the
final phase of dwarf-dwarf merging,
the total mass of new stars formed in the triggered starbursts
is significantly smaller ($\sim 4 \times 10^7 {\rm M}_{\odot}$) 
than that of old ones ($\sim 6 \times 10^8 {\rm M}_{\odot}$).
As a result of this, the central part of the merger remnant
can not be dominated by new stars.
As shown in Fig. 5,
surface density of old  stars within the central 1 kpc of the remnant
are systematically much higher than that of new ones: 
more than an order of magnitude  differences between old and new stars
at $R \sim 1$ kpc.
The low-mass density of new stars, combined with a small amount
of gas mass
($\sim 2 \times 10^7 {\rm M}_{\odot}$),
can not allow us to claim that the remnant of the gas-poor merger
in this model is classified as a BCD.
These results imply that only gas-rich dwarf-dwarf mergers
can finally evolve into BCDs.

\begin{figure}
\psfig{file=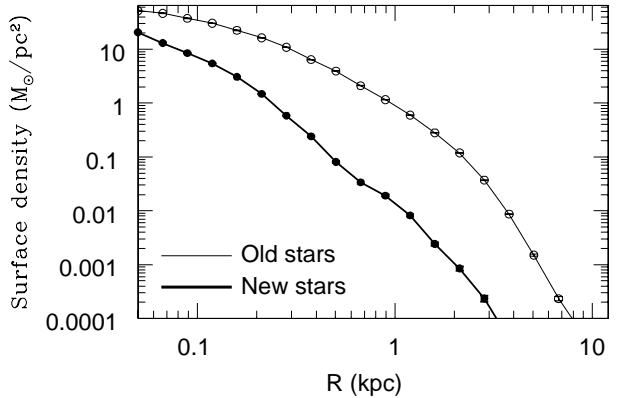,width=8.0cm}
\caption{
The  same as Fig. 2 but for the gas-poor model M4. Note that
the central part of the merger remnant is not dominated by
new stars in this model, which means that a BCD  is highly unlikely 
to be  
formed in gas-poor merging even if the merger progenitor
dwarfs have extended gas disks. 
}
\label{Figure. 5}
\end{figure}

\section{Discussion and conclusions}

The present study has  shown that 
BCDs can be mergers between dwarfs with
larger fractions of gas  
and extended gas disks. 
The present study has also shown that
(i) BCDs appear to be compact, because the compact cores
dominated by young stellar populations can be 
formed from starbursts
triggered by merging, and
(2) they also appear to be very blue, because the dominant
stellar populations are  young and  formed from
metal-poor gas during merging. 
Furthermore our models have shown  that
{\it  extended HI gas disks}  of merger progenitor dwarfs
are responsible  for the formation of BCDs with massive HI envelopes 
surrounding blue compact cores.

Recent HI observations have found that
(i) some faint galaxies have high gas mass fractions
($3<M_{\rm HI}/L_{\rm B}<27$) and (ii) fainter
galaxies are more likely to have higher $M_{\rm HI}/L_{\rm B}$
(e.g., Warren et al. 2006).
Furthermore, 
the Faint Irregular Galaxies GMRT Survey ({\it FIGGS}) has recently
revealed
that the mean value of HI-disk-to-stellar-disk ratios
($s_{\rm g}$) is about 2.4 for dIrrs
with
a median $M_{\rm B} \sim -13$ mag (Begum et al. 2008).
These observations suggest that
galaxy merging that forms  BCDs 
(i.e., mergers with high $M_{\rm HI}/L_{\rm B}$ or $f_{\rm g}$
and large $s_{\rm g}$)
can not be so rare among all types of dwarf-dwarf merging.

Although some BCDs with very metal-poor, and young stellar populations
were observationally suggested to be true young galaxies that are
currently forming (e.g., Thuan et al. 1999),
they have now being observed to have older stellar populations
(e.g., Aloisi et al. 2007).
The present study has shown that young stellar populations
in BCDs can be very metal-poor, because they originate from
extended gas disks with possibly pristine gas in merger precursor dwarfs:
the presence of very metal-poor stellar populations in BCDs  does  not 
necessarily imply the very early  epochs of their formation.

Gil de Paz et al. (2003) investigated morphological types
for  111 BCDs with H$\alpha$ emission 
and  revealed that only 10 ($\sim 9$\%) can be classified as ``iI,M''
(i.e., mergers). Our numerical results  suggest that
the ``iI'' BCDs with irregular outer halos and off-center nuclei,
which consists of 35\% of the BCDs (Gil de Paz et al. 2003),
can be formed  from dwarf-dwarf merging.
It would be  also possible that even ``iE'' BCDs with outer diffuse
elliptical halos yet without clear signs of merging are 
merger remnants with their outer halos already dynamically relaxed.
Although   ``iI C'' BCDs  with cometary morphologies 
can be  ongoing mergers with long tidal tails
viewed from edge-on,
such morphologies can be formed from other physical processes
such as ram pressure stripping.

The present simulations have shown  that evolution from BCDs formed
from dwarf-dwarf merging into gas-poor dEs is
highly unlikely owing to the presence of extended gas disks
in the BCDs.
Tajiri \& Kamaya (2002) showed that morphological evolution
from BCDs with HI envelope into gas-poor dEs 
is unlikely, because stellar feedback effects in BCDs
are not strong enough to blow off their  HI envelope.
This  observation combined with  the present numerical results
therefore implies  that BCDs are likely to evolve into gas-rich,
nucleated dIrrs 
after significant  fading of their blue compact cores.
Papaderos et al. (1996) found significant differences in
optical structures  between BCDs
and dIrrs 
and thus suggested that BCDs can evolve into dIrrs
only if BCDs can change optical structures  of their underlying
stellar populations.
By using fully consistent stellar population models combined with
chemodynamical simulations of BCD formation
in our future papers,
we discuss in detail whether the optical structures of BCDs formed from merging
can change owing to fading of their young stellar populations
so that the final structures are more similar to 
those of dIrrs
(Papaderos et al. 1996).

\section*{Acknowledgments}
I am  grateful to the anonymous referee for valuable comments,
which contribute to improve the present paper.
K.B. acknowledges the financial support of the Australian
Research Council (ARC).
The numerical simulations reported here were carried out on
SGI supercomputers 
at Australian Partnership for Advanced Computing (APAC) in Australia 
for our  research project.   
K.B. is  grateful to  Jayaram N Chengalur for his discussing
HI gas and metallicities of BCDs with me.

\end{document}